\documentclass{emulateapj}

\def\oi     {[\ion{O}{1}]}
\def\oii    {[\ion{O}{2}]}
\def\oiii   {[\ion{O}{3}]}

\def\Neiii  {[\ion{Ne}{3}]}

\def\ni    {[\ion{N}{1}]}
\def\nii    {[\ion{N}{2}]}

\def\Hi    {\ion{H}{1}}
\def\Hii    {\ion{H}{2}}

\def\aj{\rm{AJ}}                   % Astronomical Journal
\def\apj{\rm {ApJ}}                % Astrophysical Journal
\def\apjl{\rm{ApJL}}                % Astrophysical Journal, Letters
\def\apjs{\rm{ApJS}}               % Astrophysical Journal, Supplement
\def\aa{\rm{A\&A}}                % Astronomy and Astrophysics
\def\mnras{\rm{MNRAS}}             % Monthly Notices of the RAS

\usepackage{graphicx}

\usepackage{graphicx}
\begin{document}
\title{The metallicity measurement of early-type galaxies}
\author{Yu-Zhong Wu\altaffilmark{}}

\altaffiltext{}{Key Laboratory of Optical Astronomy, National
Astronomical Observatories, Chinese Academy of Sciences, Beijing
100101, China}

\shorttitle{metallicity of ETGs}
\shortauthors{Wu} \slugcomment{}

\begin{abstract}

We use data for 6048 early-type galaxies (ETGs) from Galaxy Zoo 
1 that have been cross-matched with the catalog of the MPA-JHU 
emission-line measurements for the Sloan Digital Sky Survey Data 
Release 7. We measure the metallicity of these ETGs by excluding 
various ionization sources, and study other properties as well. 
We use the optimal division line of W2–W3=2.5 as a diagnostic 
tool, and for the first time derive metallicity measurements for 
2218 ETGs. We find that these ETGs actually are closer to \Hii~ 
regions as defined by Kauffmann et al. in the 
Baldwin–Philips–Terevich diagram, and they display younger stellar 
populations. We present a full mass– metallicity relation and find 
that most ETGs have lower metallicities than star-forming galaxies 
(SFGs) at a given galaxy stellar mass. We use five metallicity 
calibrators to check our results. We find that these metallicity 
indicators (R23, O32, and O3S2) give consistent results. We 
suggest that the remaining two metallicity calibrators, which 
increase metallicity by N-enrichment, can be used to calibrate 
metallicities for SFGs, but not to estimate the metallicities of 
ETGs.

%We derive the data of 6 048 ETGs cross-matched 
%the $Galaxy~Zoo~1$ with the catalog of the SDSS DR7 MPA-JHU 
%emission-line measurements.
%We explore the metallicity measurement of ETGs by excluding
%various ionization sources, and study the properties of 
%these ETGs. We use the optimal division line of W2-W3=2.5 as the
%diagnostic tool, and derive the large sample of 
%2 218 ETGs with metallicity measurements
%for the first time. We find that these ETGs actually 
%are closer to \Hii~ regions defined by Kauffamann et al. (2003) 
%in the BPT diagaram, and they display younger stellar 
%populations. We present completely the MZ relation
%and find that most ETGs have lower metallicities than SFGs
%at a given galaxy stellar mass. 
%We use the five metallicity calibrators to check
%the result. We find that these metallicity indicators 
%(R23, O32, and O3S2) can give consistent results.
%We suggest that remaining two metallicity calibrators, which 
%metallicity increases with N rerichment, can be used to calibrate
%well metallicities for SFGs, however may not to estimate
%metallicities of ETGs.

\end{abstract}

\section{INTRODUCTION}

Early-type galaxies (ETGs) generally appear as red, 
passive/retired objects, almost without gas and dust.
lenticular and elliptical galaxies are ETGs; recent studies 
have demonstrated that ETGs usually contain a multiphase 
interstellar medium (ISM; Herpich et al. 2018) comprising 
neutral hydrogen (e.g., Krumm \& Salpeter 1979; 
Goudfrooij et al. 1994; Oosterloo et al. 2010; 
Serra et al. 2012; Lagos et al. 2014; Woods \& Gilfanov 2014), 
molecule gas (Combes, Young \& Bureau 2007; Davis et al. 2015),
halos of hot gas (Sarzi et al. 2013), and dust 
(Goudfrooij \& de Jong 1995). Gas in the ISM tend to be 
accompanied by low efficientcy star formation (SF).

Mounting evidence reveals recent or ongoing SF in some ETGs.
Yi et al. (2005) demonstrated that recent SF activities 
could be identified in approximately $15\%$ of their sample of 
160 ETGs. Using a sample of roughly 2100 Sloan Digital Sky 
Surcey (SDSS) ETGs, Kaviraj et al. (2007) have shown that 
$\sim 30 \%$ of their sample experience low-level SF, deriving 
$\sim 1-3\%$ of the total galaxy stellar mass. This implies 
that low-level SF is common in ETGs.

Much attention has been focused on explaining SF in ETGs.
One possible cause is gas accretion introduced by minor 
mergers (Kaviraj et al. 2009; Thilker et al. 2010). 
Ger\'{e}b et al. (2016) posited that an extend ($\sim$ 60 kpc)
stellar stream provides direct evidence for gas accretion, 
showing that a merger event in GASS 3505 happened in the recent 
past. Another possible cause of SF is accretion from the 
intergalatic medium (IGM). With regard to the formation of 
Hoag's Object, Finkelman et al. (2011) suggested that the core 
and \Hi ~disks of the Hoag's Object were formed at different 
evolutianary phases, and the forming of the disks was prolonged 
by `cold' accretion of pristine gas from the IGM.

Gas-phase metallicity is crucial for studying various aspects 
of the evolution of galaxies. Several attempts to obtain 
metallicity calibrations for \Hii~ regions of ETGs have been made. 
Zhang et al. (2017) studied the impact of the diffuse ionized gas 
(DIG) on metallicity measurements at kiloparsec scales with the 
data from Mapping Nearby Galaxies at APO survey. 
Kumari et al. (2019) estimated the effect of DIG contamination on 
metallicity measurements, and obtained metallicity calibrators for 
the DIG and low-ionization emission region.

To date, metallicity measurements of ETGs have only been provided 
by Athey \& Bregman (2009), Annibali et al. (2010), Bresolin (2013), 
and Griffith, Martini \& Conroy (2019); they have obtained dozens 
of oxygen abandances from optical emission lines. Here, we 
investigate the metallicity of ETGs with a large sample, and study 
the properties of these ETGs. In section 2, we provide a brief 
description of the ETG sample and our data. Excluding several 
ionization sources, we derive the metallicity of ETGs and explore 
the properties of these ETGs in Section 3. Finally, our results 
are summarized in Section 4.

\section{THE DATA}

In this work, we utilize the data of the SDSS Seventh Data 
Release (DR7; Abazajian et al. 2009). The SDSS DR7 provides
the spectra of more than one billion galaxies, covering a 
wavelength range from 3600 to 9200 {\AA}, with mean spectral 
resolution $R\sim 1800$ and assembled from $3''$ diameter
fibres. The measurements of stellar masses, emission line 
fluxes, and star formation rates (SFRs) can be obtained from 
the SDSS DR7 catalog of Max Planck Institute for 
Astrophysics$-$John Hopkins University (MPA-JHU), which is 
publically available 
\footnote{https://wwwmpa.mpa-garching.mpg.de/SDSS/DR7/}; 
this catalog supplies the spectra measurements of about 
900,000 galaxies.

\begin{figure}
\begin{center}
\includegraphics[width=8cm,height=6cm]{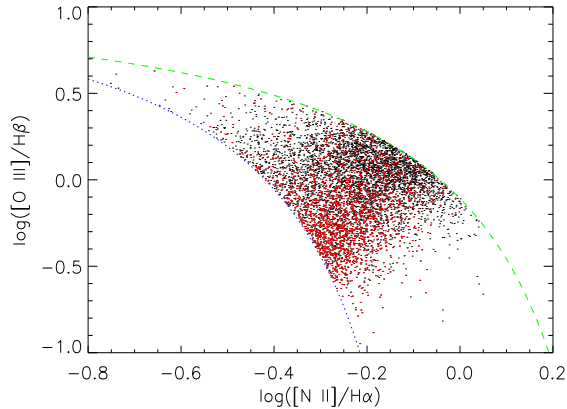}
\caption{BPT diagnostic diagram. The black and red dots
show initial composite ETGs selected; the red dots are final 
ETGs with metallicity measurements. The dotted blue curve 
represents the Kauffmann et al. (2003) semi-empirical
lower limit for the SFGs; the green dashed curve on this 
diagram is the theoretical ``extreme starburst line'' 
obtained by Kewley et al. (2001) as an upper boundary for
SFGs.}
\end{center}
\end{figure}

First, we choose galaxies at $0.04<z<0.12$ to avoid the bias 
of mass-metallicity (MZ) relations from the aperture 
(Kewley et al. 2005). Also, the aperture-covering fractions 
are required to be $>20\%$ for all galaxies, and the 
paremeters are computed from the r-band Petrosian and fiber 
magnitudes. For our sample we select galaxies with a 
signal-to-noise ratio (S/N)$>3$ for H$\alpha$ and H$\beta$, 
and with S/N $>2$ for \oii$\lambda \lambda$ 3227, 3229, 
and \nii$\lambda 6584$. Moreover, because the SFR FLAG keyword 
represents the measuring status of SFRs in the catalog, the 
keyword must be zero. As a result, we have an initial sample 
of 140,589 galaxies.

We first take the galaxies with $n_{\rm Sersic}>2.5$ to be ETGs, 
and then base our galaxy morphologies on Galaxy Zoo 1 
(Lintott et al. 2008, 2011) to match our sample with that found 
in Table 2 of Lintott et al. (2011). The S\'{e}rsic index comes 
from the New York University Value-Added Galaxy Catalog
\footnote{http://sdss.physics.nyu.edu/vagc-dr7/vagc2/sersic/}
(NYU-VAGC; Blanton et al. 2005). We cross-match the NYU-VAGC 
with our initial sample within $2''$, and obtain 133,101 
galaxies. Then we choose the galaxies with $n_{\rm Sersic}>2.5$, 
showing a sample of 42,916 galaxies. Next, we utilize Galaxy Zoo 
1 to select those galaxies with an elliptical probability that 
is higher than 0.5 (Herpich et al. 2018), and match them with 
those found in Table 2 of Lintott et al. (2011). This produces 
a sample of 16,623 ETGs.

\begin{figure}
\begin{center}
\includegraphics[width=8cm,height=6cm]{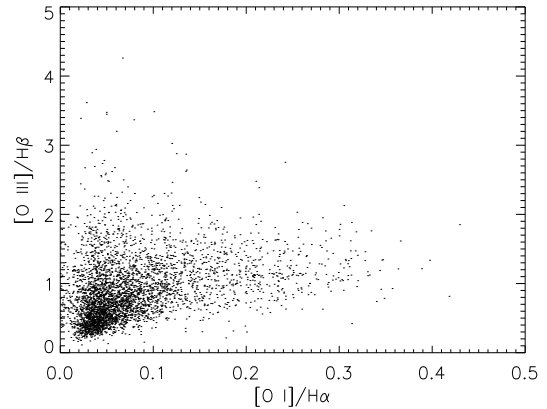}
\caption{Diagnostic diagram of \oi/H$\alpha$ and \oiii/H$\beta$,
which identifies ionization from the stellar population
with SD progenitor contribution proposed by
Woods \& Gilfanov (2014). Galaxies with \oi/H$\alpha >0.5$
are dominated by SD ionization, and exist outside of the plot.}
\end{center}
\end{figure}

We also need to assess the star-formating properties of our 
ETG sample and their probability of their contribution of an 
active galactic nucleus (AGN) on the Baldwin-Philips-Terevich 
(BPT) diagram (Baldwin et al. 1981). Based on several models, 
Kewley et al. (2001) and Kewley et al. (2006) constructed
extreme starburst lines, which are an upper limit on 
the emission-line strengths in star-forming galaxies (SFGs). 
Galaxies that lie below the two curves are dominated by star 
formation (Griffith et al. 2019). Therefore, we consider only 
those galaxies that are located in the composite region on 
the BPT diagram in this work, which means that our sample 
contains 6048 composite ETGs (see the BPT diagram in 
Figure 1).

To further study the properties of ETGs, we utilize 
the Wide-field Infrared Survey Explorer (WISE; 
Wright et al. 2010) catalog to explore their properties. 
This survey provides coverage of the whole sky in four 
bands: W1, W2, W3, and W4, with the central wavelengths 
of 3.4, 4.6, 12, and 22 $\mu m$. The accurate position,
four-band fluxes, four-band instrumental profile-fit phometry
magnitudes, and their instrucmental profile-fit photometry 
S/N ratios are publicly available 
\footnote{https://irsa.ipac.caltech.edu/cgi-bin/Gator/nph-dd}.
We cross-match the ETG sample using the ALLWISE source catalog 
within $2''$ and S/N$>3$ for W2 and W3, and obtain 4177 galaxy 
sample.

In the MPA-JHU catalog for the SDSS DR7, $M_{\star}$ and SFRs 
measurements assumed a Kroupa (2001) initial mass function (IMF) 
and are corrected by a Chabrier (2003) IMF. In SFGs, gas-phase 
oxygen abundances are estimated by using abundance measurements 
of extragalactic \Hii. Because ETGs include many other ionization 
sources, such as AGN activities, cosmic rays, shocks, and old, 
hot stars, Griffith et al (2019) obtained the metallicities
of three ETGs by considering the influence of these ionization 
sources on the abundance measurements.
In addition, Brown et al. (2016) used a sample of about 200,000 
SFG sample to estimate the performance of some abundance 
indicators, and suggested that the O3N2 method of 
Pettini \& Pagel (2004, PP04) is the most perfect calibrator. 
In the higher stellar masses, the MZ relation calibrated by the 
PP04-O3N2 indicator can display a consistency with the direct 
method ($T_e$ method) MZ relation (Andrews \& Martini 2013). 
In addition, the O3N2 indicator adopts flux ratios of more 
neighboring lines than other indicators, avoiding more 
substantial effects of atmospheric dispersion 
(Griffith et al. 2019). In this Letter,
we utilize PP04-O3N2 as our abundance estimator.

\begin{figure}
\begin{center}
\includegraphics[width=8cm,height=6cm]{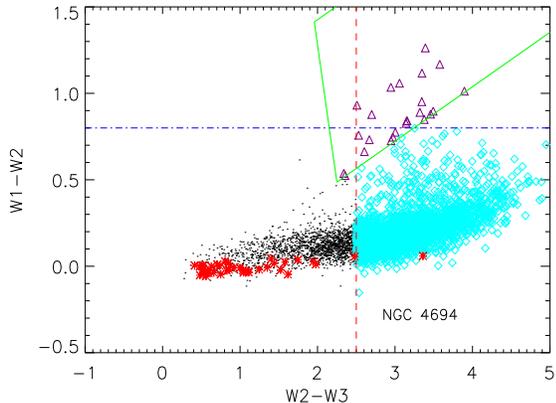}
\caption{W2-W3 vs. W1-W2 color-color diagram for composite 
ETGs. The blue horizontal dotted-dashed line represents the 
mid-IR criterion to select AGNs proposed by Stern et al. (2012).
The green solid lines display the ``AGN'' wedge as suggested
by Mateos et al. (2012). The red vertical dashed line describes
the best boundary between retired galaxies and ongoing SF 
galaxies. The black dots and purple triangles are composite ETGs, and
the former ones have no SF or AGNs, while the latter ones 
exhibit AGN activities. The cyan diamonds
are our study sample of ETGs with SF. The red
``$*$'' signs are those ETGs with metallicity measurements from
Bregman (2009), Annibali et al. (2010), Bresolin 2013, and
Griffith et al. (2019).}
%{\noindent  \vglue 0.5cm {\sc  }}
\end{center}
\end{figure}

In addition to the metallicity calibrator of PP04-O3N2, in 
this Letter we also use another five metallicity calibrators to 
calculate oxygen abundance of ETGs; Tremonti et al. 2004 (T04), 
Jones et al. (2015, Jon15), Curti et al. (2017, Curti17), 
S\`{a}nchez-Almeida et al. (2018, S\'{a}nch18), and
Sanders et al. (2018, Sander18).

\section{Results}

\subsection{The Metallicity Measurement of ETGs}

The gas-phase oxygen abundance of the warm ISM can be estimated 
by metallicity measurements of extragalatic \Hii~ regions and 
photoionization models. In ETGs, many ionization sources, such 
as AGN activities, shocks, post-asymptotic giant brance (PAGB) 
stars, cosmic rays. In this work, the AGN activities can be 
excluded due to the fact that our sample consists of composite 
ETGs, dominated by SF (Griffith et al. 2019),
and we do not need to consider the AGN photoionization. Regarding
the shock excitation mechanism, the observationed line fluxes
can be influenced by shocks. Compared with the typical densities 
and velocities suggested by Sparks et al. (1989), the shock
energy is too low by two orders of magnitude 
(Athey \& Bregman 2009). In Griffith et al. (2019), we can see 
that \nii/H$\alpha$ ratio excited by the shock from a Small 
Magellanic Cloud-like galaxy (Allen et al. 2008) is far lower 
(about one order of magnitude) than the ratio of galaxies lying 
in composite or AGN regions. From Figure 4 of Griffith et al. 
(2019), we can also see that the line flux ratio excited by the 
cosmic ray or extra heat is far lower than that excited by 
galaxies located in composite or AGN region on the BPT diagram, 
and excludes the two ionization sources. In addition, because 
PAGB stars adopting the excitation source can provide energy on 
the same order of magnitude as weak AGNs, Belfiore et al. (2016) 
suggested that many galaxies, dominated by SF, are misclassified 
as active ones (Griffith et al. 2019).

\begin{figure*}
\begin{center}
\includegraphics[width=16cm,height=12cm]{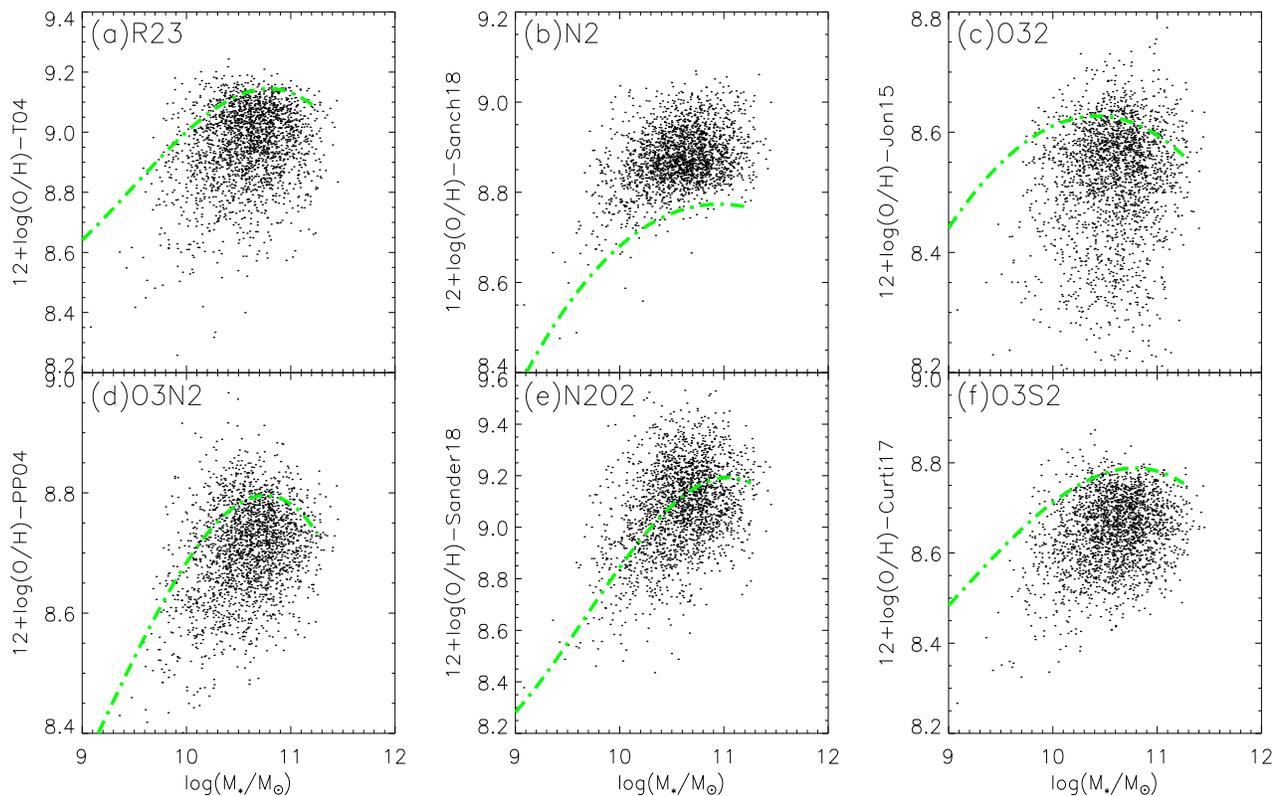}
\caption{Comparison of the MZ relations of composite ETGs
for different metallicity calibrators. 
The green dotted-dashed lines represent a polynomial fit 
of data from 93,089 SFGs (see the text).}
\end{center}
\end{figure*}

Because the probability of the \Neiii $\lambda 3869$ 
photoionization transition is far lower than 
one of the collisional cross-section, Athey \& Bregman (2009) 
suggested that the line is an important index of collisinal 
excitation; they excluded NGC 4125 and NGC 2768 from their 
sample due to the fact that they have clear 
\Neiii $\lambda 3869$ fluxes. This demonstrates that the two 
objects have more a complicated excitation mechanism than 
photoionization. We exclude two objects having significant 
\Neiii $\lambda 3869$ line, and obtain 4175 ETGs.

Griffith et al. (2019) introduced a dianostic diagram of 
\oi $\lambda 6300$/H$\alpha$ versus 
\oiii $\lambda 5007$/H$\beta$ to exclude an ionization mechanism, 
which is excited by single-degenerate (SD) SNe Ia progenitors 
(Woods \& Gilfanov 2014). In this model, Woods \& Gilfanov (2014) 
employed diagnostic tools, utilizing \oi, \ni~ emission lines 
increasingly ionized by high-temperature, to test if WD 
progenitors could be a dominant ionization sourece. In Figure 2, 
we show our sample on the diagnostic diagram of 
\oi $\lambda 6300$/H$\alpha$ versus 
\oiii $\lambda 5007$/H$\beta$, obtaining 4097 ETGs.

Figure 3 presents the relation between W2-W3 and W1-W2 colors 
for composite ETGs. The optimal division line (red dashed line), 
W2-W3=2.5, is used as the best demarcation between galaxies 
with and without SF (Herpich et al. 2016). 
Stern et al. (2012) proposed that W1-W2$>0.8$ of Figure 2 be 
the mid-IR standard by which to choose AGNs, and these 
measurements reveal that nuclear activity provides almost 
all the IR emission (Caccianiga et al. 2015). The `AGN wedge' 
proposed by Mateos et al. (2012) is displayed by the green 
lines in Figure 2.

In Figure 3, the black dots at the left of the red dashed line 
W2-W3=2.5 show 1845 composite ETGs without SF. The red `$*$' 
signs in Figure 3 represent those ETGs with metallicity 
measurements from Athey \& Bregman (2009), 
Annibali et al. (2010), Bresolin 2013, and 
Griffith et al. (2019). One of the galaxies at the right of the 
red dashed line of Figure 3 is NGC 4694, which appears in the SF 
region of the BPT diagram of Figure 4 of Griffith et al. (2019).
This galaxy should be an SF ETG. The purple triangles locate in 
the `AGN' wedge (green solid lines) proposed by 
Mateos et al. (2012) are used to display ETGs with nuclear 
activities; here there are 24 such ETGs. The cyan diamonds, which 
lie to the right of this red dashed line of Figure 3, are thought 
to be composite ETGs with SF, because Herpich et al. (2016) found 
that galaxies with mid-IR color W2-W3$>2.5$ denote objects with 
SF. Less than half of all composite ETGs have SF, and we suggest 
that SF is a dominant excitation source in these ETGs. Finally, 
we derive our sample of 2218 ETGs.

\begin{figure}
\begin{center}
\includegraphics[width=8cm,height=6cm]{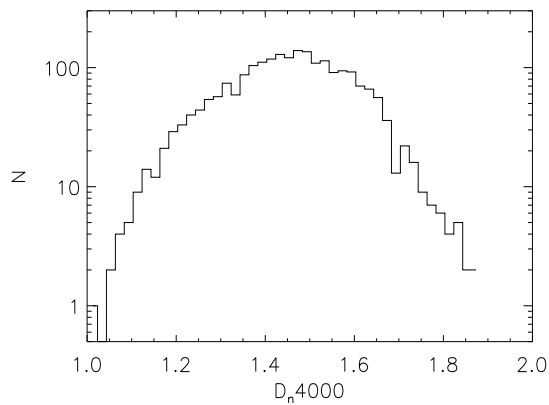}
\caption{Distribution of $D_{n}4000$ for our ETG sample
with metallicity measurements.}
%{\noindent  \vglue 0.5cm {\sc  }}
\end{center}
\end{figure}

\subsection{Properties of these ETGs}

In Figure 1, we use the BPT diagnostic diagram to show the 
ETG samples. The blue dotted curve represents the
Kauffmann et al. (2003) semiempirical lower boundary for 
SFGs, while the green dashed curve on this diagram show the 
theoretical ``extreme starburst line'' obtained by 
Kewley et al. (2001) as an upper boundary for SFGs. The 
black and red dots present initial composite ETGs; the red 
dots display our final ETG sample with metallicity measurements. 
From Figure 1, we can see that most of the red dots are 
close to the blue dotted curve, and a majority of the black 
dots are close to the green dashed curve. This indicates that 
the ETGs represented by red dots may be dominated by SF.

Due to the limited sample size of ETGs, we were previously 
unable to completely present their MZ relation. Here, we have a 
chance to present it using a large ETG sample. Figure 4 shows 
the MZ relation for our composite ETGs with the six metallicity 
calibrators. We find that the range of most galaxy stellar masses 
is from log$(M_{\star}/M_{\sun}) \sim 10.0$ to 
log$(M_{\star}/M_{\sun}) \sim 11.0$. We find that these MZ 
relations calibrated by the six metallicity indicators
do not show clearly positive correlations, and that most ETGs 
lie below the green fit curves in Figure 4. Next we discuss the 
properties of the MZ relation in the following two cases: the 
MZ relation of ETGs that lie below and above the median MZ 
relation of SFGs.

From Figures 4(a), 4(c), 4(d), and 4(f), the MZ relations for ETGs
lie below the median MZ relations for SFGs. They employ, respectively, 
the metallicity indicators of R23, O32, O3N2, and O3S2 to calibrate 
the ETG metallicities. In Figure 4(a), the green dotted-dashed line 
describes a polynomial fit of data (deriving 93089 SFGs from the 
MPA-JHU catalog for the SDSS DR7 with the same sample selection method 
of Wu et al. (2019), except S/N $>2$ for \oii$\lambda \lambda$ 3227, 3229, 
and \nii$\lambda 6584$.), which have median values of 30 bins, 0.1 dex 
in mass, and include more than 100 galaxies. The MZ relation does not 
clearly show a positive correlation, with the Spearman coefficient r=0.17.
Also, the MZ relations calibrated by metallicity indicators of Jon15, 
PP04, and Curti17 do not present a positive correlation, with the 
Spearman coefficient r=0.1, r=0.26, and 0.21, respectively. We can see 
that most of the composite ETGs displayed in Figure 4(a) lie below the 
green fit line at a given stellar mass. Figures 4(c), 4(d), and 4(f) also 
show the same result, with a majority of ETGs lying below the median MZ 
relation of SFGs. The middle panel of Figure 9 of Griffith et al. (2019) 
shows the MZ relation of ETGs using the O3N2 metallicity 
calibration of Pettini \& Pagel (2004). We can see that most ETGs may be 
below the MZ relation of SFGs. Combined the results of our MZ relation and 
that of Griffith et al. (2019), we find that the MZ relation of these ETGs 
may be different from that of late-type galaxies; this indicates that 
the metallicities of ETGs are likely to be lower than thoss of late-type 
galaxies at a fixed stellar mass.

Figures 4(b) and 4(e) do not show a significantly positive correlation,
with the Spearman coefficient r=0.28 and r=0.31, respectively.
In these two Figures we find that the metallicity calibrators 
of S\'{a}nch18 and Sander18 do not show the result of Figures 4(a), (c), 
(d), and (f); in addition, almost all fo theETGs lying above the green 
fit line in adopting N2 metallicity indicator.
In Figure 4(b), The S\'{a}nch18 calibrator employs the N2 indicator 
(\nii$\lambda 6584$/$\rm H \alpha$) and Equation (3) of 
S\'{a}nchez-Almeida et al. (2018). Their metallicity increases 
with increasing N2. Similarly, in Figure 4(e), we find that about 
$52\%$ of ETGs lie below the green fit curve. We suggest that this may 
originate from the metallicity calibrator of Sander18, and the
metallicity increases with increasing indicator 
N2O2 (\nii $\lambda 6584$/(\oii$ \lambda \lambda3727, 3729$)).
In the top panel of Figure 9 of Griffith et al. (2019), we find that 
most ETGs from Athey \& Bregman (2009), Annibali et al. (2010), and 
Griffith et al. (2019) have higher, about 0.2 dex, than SFGs in 
O3N2 metallicity calibrated from the PP04-N2 metallicity indicator.
In the bottom panel, we also find that these ETGs have higher 
about 0.1 dex in metallicity than SFGs using PP04-O3N2 indicator 
calibrated from the KD02-N2O2 estimator. We find that their results 
are almost consistent with our results shown in Figures 4(b) and 4(d).
According to Coziol et al. (1999) model, the nitrogen production 
occures almost ceaselessly in SFGs, but the oxygen enrichment is only
produced in massive star ecolution, and the two productions
proceed alternately (Wu \& Zhang 2013). It is difficult to observe 
oxygen production in ETGs, while nitrogen production is easier to 
observe because ETGs tend to experience low-level SF (Yi et al. 2005;
Kaviraj et al. 2007). Therefore, we suggest that these metallicity
calibrators, wherein metallicity increases with N-enrichment, for 
example, the S\'{a}nch18 and Sander18 metallicity indicators, can be 
used to successfully calibrate metallicities for SFGs, but may not 
to estimate the metallicities of ETGs.

In Figure 5, we show the distribution of $D_{n}4000$ for our final 
ETG sample. We find that the majority of ETGs with metallicity 
measurements have younger stellar populations, and about $85\%$ of 
these ETGs display $D_{n}4000 < 1.6$. These ETGs with W2-W3$ >2.5$
clearly show SF (Herpich et al. 2016), and approach the 
Kauffmann et al. (2003) semiempirical line; therefore, most of them 
should have younger stellar populations.

\section{Summary}

In this Letter we derive the observational data of 6048 ETGs, and 
cross-match Galaxy Zoo 1 data with SDSS DR7 MPA-JHU emission-line 
measurements. We exclude various ionization sources to explore the 
metallicity measurement of ETGs, and investigate the properties of 
these ETGs. We summarize our main results as follows:

1. We utilize the optimal division line of W2-W3=2.5 as the diagnose 
tool, which selects ETGs (W2-W3$>2.5$) with SF as our composite ETG 
sample. We obtain the final sample of 2 218 ETGs, and can calculate 
their metallicities.

2. We find that our ETGs tend to be located close to the semiempirical 
lower limit for SFGs proposed by Kauffmann et al. (2003) in the BPT 
diagram. Moreover, we find that these ETGs have younger stellar 
populations, and $85\%$ of the ETGs have $D_{n}4000 < 1.6$.

3. The MZ relation for ETGs is shown, and we find that the metallicity 
of these ETGs tends to be lower than SFGs at the same galaxy stellar mass. 
We use five metallicity calibrators to check the result. We find that 
three of these metallicity indicators (R23, O32, and O3S2) can give 
consistent results, while the N2O2 and N2 indicators do not obtain the 
same result. We suggest that these metallicity calibrators, wherein 
metallicity increases with N-rerichment (for instance, S\'{a}nch18
and Sander18 metallicity indicators), can be used to successfully 
calibrate metallicities for SFGs, but may not be able to estimate
the metallicities of ETGs.

\acknowledgments

Y.Z. Wu thanks the anonymous referee for valuable suggestions and comments that improved the quality of this Letter. This work was supported by the Natural Science Foundation of China (NSFC; No. 11703044).

\end{document}